\documentclass [superscriptaddress,reprint,amsmath,amssymb,aps,prl,showkeys,floatfix]{revtex4-2}
\usepackage{xcolor}
\usepackage{soul}
\usepackage[normalem]{ulem}
\usepackage[hidelinks]{hyperref}
\usepackage{graphicx}
\usepackage{dcolumn}
\usepackage{bm}
\usepackage{color}
\usepackage{float}
\usepackage[utf8]{inputenc}
\usepackage[mathlines]{lineno}
\usepackage{multirow}
\usepackage{amsmath}
\usepackage{pdfpages}
\usepackage{pgffor}
\makeatletter
\AtBeginDocument{\let\LS@rot\@undefined}
\makeatother
\usepackage{hyperref}
\usepackage{soul,xcolor}
\usepackage{natbib}
\usepackage{amsmath} 
\usepackage{xcolor}
\usepackage{xr}
\usepackage{chemformula}
\usepackage{siunitx}
\usepackage{physics}
\soulregister\cite7
\soulregister\ref7
\setstcolor{blue}
\newcommand{\doHMN}[2]{%
  \begingroup\lccode`~=`#1
  \lowercase{\endgroup\let~}#2%
  \mathcode`#1="8000
}

\makeatletter
\newcommand*{\addFileDependency}[1]{
  \typeout{(#1)}
  \@addtofilelist{#1}
  \IfFileExists{#1}{}{\typeout{No file #1.}}
}
\makeatother

\usepackage{tikz}

\begin{document}
\preprint{APS/123-QED}
\title{Large photo-induced tuning of ferroelectricity in sliding ferroelectrics}
\author{Lingyuan Gao}
\affiliation{Smart Ferroic Materials Center, Physics Department and Institute for Nanoscience and Engineering, University of Arkansas, Fayetteville, Arkansas, 72701, USA}
\author{Laurent Bellaiche}
\email{laurent@uark.edu}
\affiliation{Smart Ferroic Materials Center, Physics Department and Institute for Nanoscience and Engineering, University of Arkansas, Fayetteville, Arkansas, 72701, USA}

\date{\today}

\begin{abstract}
Stacking nonpolar, monolayer materials has emerged as an effective strategy to harvest ferroelectricity in two-dimensional (2D) van de Waals (vdW) materials. At a particular stacking sequence, interlayer charge transfer allows for the generation of out-of-plane dipole components, and the polarization magnitude and direction can be altered by an interlayer sliding. In this work, we use {\it ab initio} calculations and demonstrate that in prototype sliding ferroelectrics 3R-stacked bilayer transition metal dichalcogenides MoS$_2$, the out-of-plane electric polarization can be robustly tuned by photoexcitation in a large range for a given sliding. Such tuning is associated with both a structural origin---i.e., photoinduced structural distortion, and a charge origin---namely, the distribution of photoexcited carriers. We elucidate different roles that photoexcitation plays in modulating sliding ferroelectricity under different light intensities, and we highlight the pivotal role of light in manipulating polarization of 2D vdW materials.  
\end{abstract}

\maketitle
Ferroelectric materials inherently possess an electric polarization, which can be reoriented by external electric field. Leveraging on this fascinating property, many devices such as random-access-memory, actuators, detectors and modulators, have been successfully developed based on ferroelectrics~\cite{lines2001principles}. While conventional ferroelectrics are primarily bulk oxides with strong covalent and ionic bonds, many recent studies revealed emergent ferroelectricity discovered in 2D vdW materials~\cite{wang2023towards,zhang2023ferroelectric,man2023ferroic,fan2023recent}. Compared to conventional ferroelectrics, vdW ferroelectrics demonstrate strong resistance to depolarization fields at ultrathin thickness, presenting them as great candidates for electronics beyond Moore's law; the weak vdW interaction between layers also facilitates easy mechanical exfoliation of layered materials. Experimentally verified 2D vdW ferroelectric systems include but are not limited to In$_2$Se$_3$ with interlocked polarization along different directions~\cite{ding2017prediction,zhou2017out}, CuInP$_2$S$_6$ with quadruple potential wells for ion displacements~\cite{liu2016room,brehm2020tunable}, polar metal WTe$_2$~\cite{fei2018ferroelectric,yang2018origin}, and monolayer group-IV monochalcogenides SnS/Te/Se~\cite{chang2016discovery,barraza2021colloquium}.     

Given notable 2D ferroelectrics listed above, native 2D ferroelectric materials are still rare due to constraints imposed by polar symmetry groups. In 2017, a theoretical proposal of stacking bilayers or multilayers in a noncentrosymmetric order opens up a new route towards building 2D ferroelectrics from non-polar parent materials~\cite{li2017binary,wu2021sliding}. By applying a lateral shift to one layer relative to the other in a bilayer system, two stable configurations linked by mirror symmetry with  respect to horizontal plane can be converted interchangeably. As a result, the vertical polarization is switched by this shift at a very low energy cost. In following years, experiments have confirmed this theoretical prediction in several 2D bilayer vdW systems, including AB/BA stacking boron nitride (BN)~\cite{yasuda2021stacking,vizner2021interfacial}, and transition metal dichalcogenides (TMD) with 3R stacking~\cite{wang2022interfacial,weston2022interfacial,wan2022room, rogee2022ferroelectricity}. Sliding is also applied to multilayer stacked systems, where multiple polarization states are successfully created~\cite{meng2022sliding,deb2022cumulative}. Unlike conventional ferroelectrics where polarization arises from displaced ions off highly symmetric positions, here the ferroelectric origin is ``electronic": the polarization develops from the interlayer charge transfer between two weakly coupled layers.  Nevertheless, the unique origin also limits the polarization magnitude to small values~\cite{wu2021sliding}. Thus, it is crucial to explore potential strategies for engineering sliding ferroelectrics to harness a larger polarization. 

Various approaches involving applying external stimuli have been explored to engineer 2D ferroelectricity. For example, strain engineering~\cite{chen2022large,dong2023giant} and chemical doping~\cite{xue2022discovery,sui2023sliding} have demonstrated the effectiveness of switching or enhancing polarization. With advancements in laser techonology, light has become another prevalent tool in manipulating ferroelectricity: terahertz laser have been employed to drive optical phonons and change polarization~\cite{chen2016ultrafast,mankowsky2017ultrafast,nova2019metastable,li2019terahertz}.In addition, ultrafast light pulses with above-bandgap photoexcitation have also been used to control polarization via altering the charge order and electronic phases~\cite{jiang2016origin,iwano2017ultrafast,porer2018ultrafast,lee2021structural,yu2024ultrafast,koshihara1990photoinduced,perfetti2006time,rohwer2011collapse,zhang2016cooperative,kogar2020light}; however, their effect on sliding ferroelectricity, which is based on an electronic order, remains unknown. 
\begin{figure}
\includegraphics[width = 80mm]{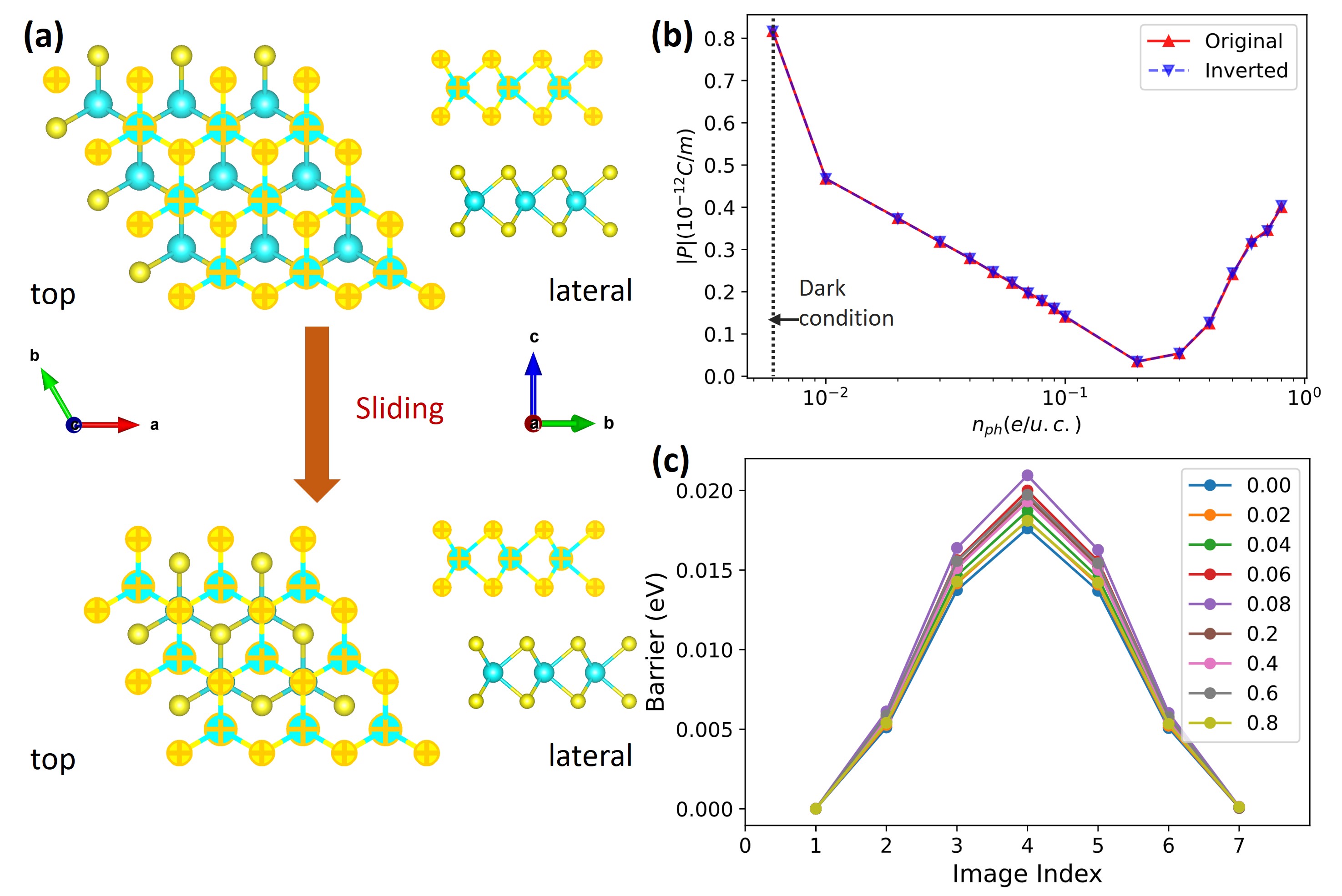}
\caption{(a) Top and lateral views of two  3R-stacked bilayer MoS$_2$ structures, which can be interconverted either by in-plane sliding or an out-of-plane mirror inversion. The blue and yellow balls denote Mo and S atoms, respectively. (b) Polarization magnitudes of out-of-plane component $P_z$ at different numbers of photoexcited carriers $n_{ph}$ in the upper (``original") and lower (``inverted") structure of (a), respectively. $P_z$ in the ``inverted" structure has an opposite sign to $P_z$ in the ``original" structure. (c) Energy pathways for converting two structures at different $n_{ph}$. The number on horizontal axis represents the image index of each optimized intermediate configuration/image along the transition pathway.}
\end{figure}

In this Letter, via {\it ab initio} calculations, we study how photoexcitation can influence sliding ferroelectricity of 2D vdW materials. Rhombohedrally(3R)-stacked bilayer MoS$_2$ is selected as the model system. We anticipate that the interplay between photoexcited carriers and intrinsic electrons can adjust the electron-dominated polarization in a more delicate manner. By varying the light fluence in a wide range, we uncover an intriguing effect of photoexcitation on sliding ferrelectricity:  photoexcitation significantly screens the intrinsic polarization at low fluence level, whereas additional photoexcited carriers introduced at higher fluence levels aid in the recovery of the diminishing polarization. Such nontrivial, non-monotonic trend is distinct from the conventional ferroelectric oxides, where polarization tends to vanish and non-polar phase is favored with an increasing illumination~\cite{paillard2016photostriction,paillard2019photoinduced}. Furthermore, when light fluence reaches a critical value, it induces a structural phase transition from $P3m1$ to $Cm$ phase, accompanied by the softening of $M$-point phonon modes. This finding, along with other recently discovered novel photoinduced phases in monochalcogenides~\cite{matsubara2016initial,huang2022observation,mocatti2023light,dringoli2024ultrafast,furci2024first}, indicates that phase transition inaccessible by thermal activation can be induced by photoexcitation~\cite{de2021colloquium,rajpurohit2022photo, bargheer2004coherent,fritz2007ultrafast,beaud2014time,wall2018ultrafast}. Compared to the ground-state structure, the transient $Cm$ phase exhibits a polarization with a magnitude enhanced by over 5 times, demonstrating that photoexcitation can effectively enhance the functionality of sliding ferroelectrics. We attribute the effect to a structural origin related to ionic displacements under light and a charge origin related to the distribution of photoexcited carriers. 

Technically, we perform constrained density functional theory (DFT) calculations implemented in the Quantum Espresso software package to simulate the ultrafast process under light~\cite{giannozzi2009quantum,giannozzi2017advanced,marini2021lattice}. Within this approximation, two separate chemical potentials representing the thermalization of electrons and holes respectively are established and remain fixed throughout the self-consistent calculation, resembling a thermalized electron-hole plasma. Such approximation can well describe the transient process when pumped electrons and holes populate conduction and valence bands respectively right after the photoillumination~\cite{tangney1999calculations,tangney2002density,murray2005effect,murray2007phonon, fritz2007ultrafast,paillard2016photostriction,haleoot2017photostrictive,porer2018ultrafast,paillard2019photoinduced,porer2019ultrafast,tian2019optically,gu2021carrier,mocatti2023light,peng2024photoinduced}.
Since position operator is not well defined in infinitely expanded crystals~\cite{blount1962formalisms}, Berry phase (BP) method is commonly used to compute the electric polarization of a periodic system~\cite{king1993theory, vanderbilt1993electric, resta1994macroscopic}. Here, as the bilayer system is confined along the out-of-plane direction, we can directly compute the out-of-plane component of polarization in the classical version using charge density integration; this also allows us to circumvent the need for the system to be insulating, as required by the BP method~\cite{king1993theory, vanderbilt1993electric, resta1994macroscopic}. Benchmark calculations on different systems show good agreement between{ polarizations computed by BP method and by integration of charge density, and they are given in Supplementary Material (SM) along with other computational details~\cite{supplementarymaterial}.

It is well known that two-dimensional TMDs can host many-body effects, where electrons and holes are tightly bound due to the strong attractive Coulomb interaction~\cite{mueller2018exciton, wang2018colloquium, regan2022emerging}. As such, we model the scenario of strong light illumination, where the attractive Coulomb interaction is strongly reduced by the screening of high density of carriers, and a dense electron-hole plasma is created~\cite{chernikov2015electrical,chernikov2015population,wang2019optical}; under this condition, the single-particle approximation remains valid. Given that critical carrier densities for reaching electron-hole plasma in TMDs are estimated to be $\sim 10^{13}$ cm$^{-2}$~\cite{chernikov2015electrical,chernikov2015population,steinhoff2017exciton,lin2019many,bataller2019dense}, we begin by introducing thermalized electron-hole (e-h) pairs at a density of 0.01$ \ e /$unit cell (u.c.). In constrained DFT, the density of photoexcited carriers $n_{\rm{ph}}$ is approximated to be linearly proportional to the light fluence~\cite{marini2021light,paillard2023light,gao2023photoinduced,peng2024photoinduced}, and corresponding fluences are given in SM. The main $n_{\rm{ph}}$ studied in this work ranges from 0.1 to $1.0 \ e /$u.c., and that corresponds to a carrier density of $10^{14}\sim 10^{15}$ cm$^{-2}$ and an estimated fluence of 1$\sim$10 mJ/cm$^2$. As studied in previous experiments, for fluence at this order, heating effect on MoS$_2$ is limited while the electronic effect is dominant~\cite{paradisanos2014intense,mannebach2014ultrafast,khatua2022ultrafast}; also, high-density electron-hole plasma at $\sim 10^{14} /$cm$^2$ has been successfully created in bilayer TMD with photoexcitation~\cite{chernikov2015electrical,wang2019optical}.

Figure 1(a) presents two noncentrosymmetric, 3R-stacked bilayer MoS$_2$ structures: in the upper (``original'') structures, the top layer's sulfur atoms sit at the hexagon's center from the top view, while in the lower (``inverted'') structures, the top layer's molybdenum atoms sit at the hexagon's center. One structure can be converted to the other either by shifting the bottom layer along $\frac{1}{3} (\Vec{a} - \Vec{b})$, or by a mirror flip relative to the $\bm{ab}$ plane from the lateral view, demonstrating that out-of-plane polarization $P_z$ are opposite in these two structures. 
\begin{figure}
\includegraphics[width = 90mm]{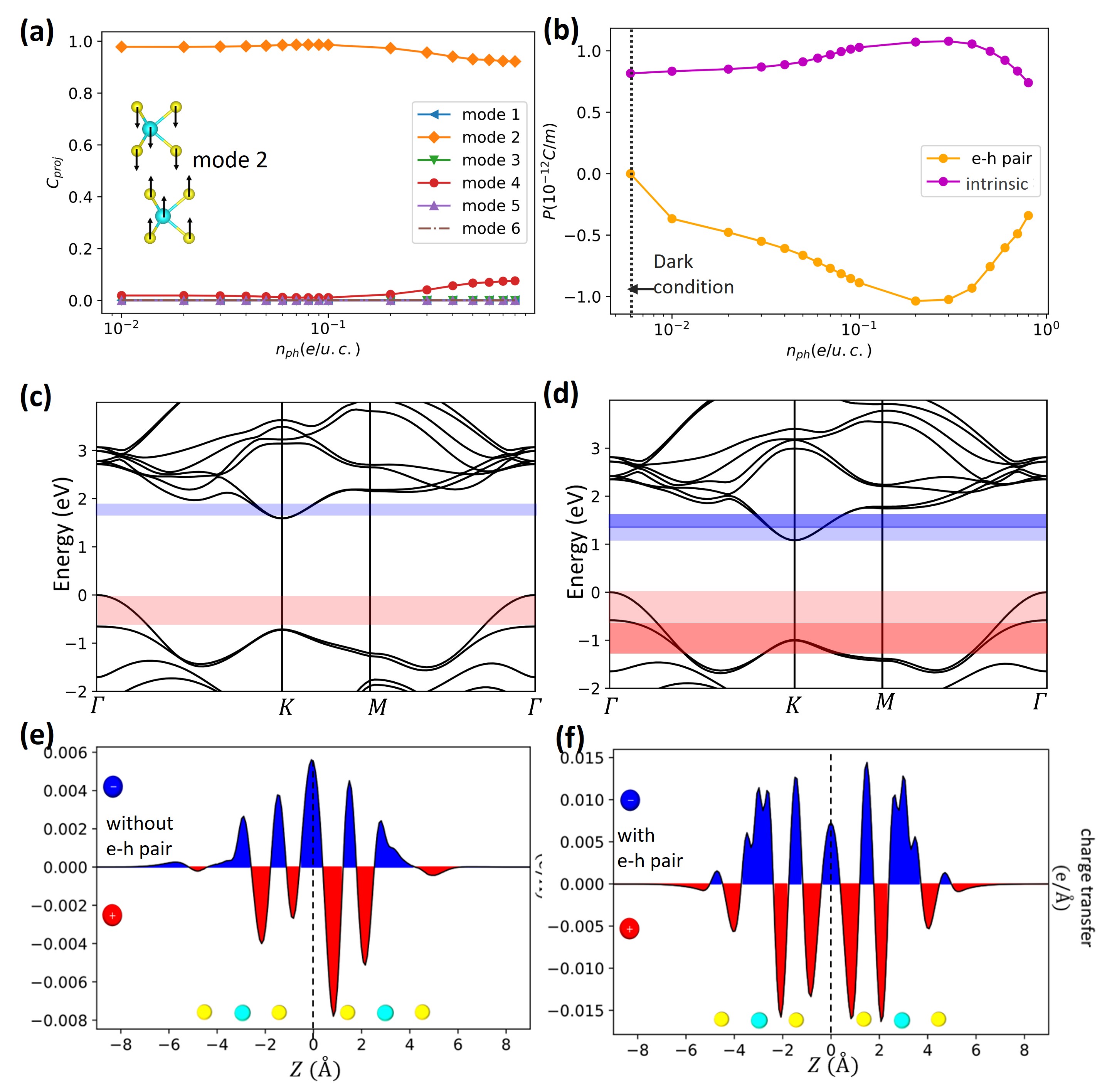}
\caption{ (a) Projection of out-of-plane ion displacements on 6 $A_1$ phonon modes at different $n_{\rm{ph}}$. (b) The structure- and charge-contributed components  of $P_z$ at different $n_{\rm{ph}}$, which are associated with the relaxed structures under photoexcitation, and photoexcited e-h pairs respectively. (c)(d) Electronic band structure at $n_{\rm{ph}} = 0.2 \ e/$u.c. and $0.8 \ e/$u.c., respectively. Regions occupied by excited electrons and excited holes are highlighted by blue and red, respectively. (e)(f) Charge density difference $\rho_{\rm{CDD}}$ \ (or charge transfer) from independent monolayer to stacked bilayer MoS$_2$ at dark condition and at $n_{\rm{ph}} = 0.2 \ e/$u.c. respectively. Blue (positive) denotes the electron accumulation and red (negative) denotes the electron depletion. }
\end{figure}

To mimic the light effect, full structural relaxations are performed first following the injection of $n_{\rm{ph}}$ e-h pairs into bilayer systems, and $P_z$ are calculated subsequently using self-consistent charge densities. The variation of $P_z$ with $n_{\rm{ph}}$ for both structures are shown in Fig. 1(b). Consistent with structural analysis above, $P_z$ in ``original" and ``inverted"" structures have the same magnitude but different signs. The effect of thermalized carriers on polarization is quite notable: when $n_{\rm{ph}} = 0.01 \ e/$u.c. ($\sim 10^{13} \ \rm{cm}^{-2}$), $P_z$ is already reduced from $0.82 \ \rm{pC}/\rm{m}$ at dark condition to $0.47 \ \rm{pC}/\rm{m}$, by over 40$\%$. The polarization keeps decreasing with an increasing $n_{\rm{ph}}$, and it reaches the minimal value of $0.03 \ \rm{pC}/\rm{m}$ when $n_{\rm{ph}} = 0.2 \ e/$u.c., indicating that the polarization is almost fully suppressed! Such $n_{\rm{ph}}$ corresponds to an estimated fluence of 1.8 $\rm{mJ}/\rm{cm}^{-2}$, which is attainable in experiment~\cite{mannebach2014ultrafast,chernikov2015population,wang2019optical}. Nevertheless, if we further increase $n_{\rm{ph}}$, $P_z$ displays an upward trend. Although the increase is not as pronounced as the initial decrease at smaller $n_{\rm{ph}}$, it does recover to $0.4 \ \rm{pC}/\rm{m}$ when $n_{\rm{ph}} = 0.8 \ e/$u.c.. As a comparison, for doping introduced by electrostatic gating, it is reported that the carrier density at $10^{13} \  \rm{cm^{-2}}$ leads to a $\sim$40$\%$ reduction on polarization~\cite{deb2022cumulative}, while another study shows that  the ferroelectric polarization remains robust when carrier density is as high as  $\sim4 \times 10^{14} \ \rm{cm^{-2}}$~\cite{meng2022sliding}. Translation pathways for converting these two structures at different $n_{\rm{ph}}$ are given in Fig. 1(c), and low energy barriers varying between $17\sim 21$ meV/u.c denote small energy costs for switching polarization under light, or ``photodoping". Note that the minimal energy barrier $\approx$ 17 meV occurs at dark condition $n_{\rm{ph}} =  0 \ e /$u.c.. 

The change in polarization can typically be attributed to either an ionic or electronic origin. Here, we consider the ionic effect driven by light first. Below we only focus on the ``original" structure as it can be easily mapped to the ``inverted" structure. Following an ultrafast above-bandgap excitation, a rapid change of carrier distribution reshapes potential energy surfaces, leading to ion displacements from their original equilibrium positions. This is known as ``displacive excitation of coherent phonons" (DECP) mechanism, where ions move coherently along the coordinate of a fully symmetric $A_1$ phonon mode~\cite{zeiger1992theory,kuznetsov1994theory,hunsche1995impulsive,tangney1999calculations,lakehal2019microscopic,gu2021carrier,caruso2023quantum}. 3R bilayer MoS$_2$ has 18 $\Gamma$-point phonon modes that can be decomposed into $6A_1 + 6E$ modes~\cite{park2019ferroelectric}, and all $A_1$ modes are motions along the out-of-plane $z$ direction. As shown in Fig. 2(a), by projecting $z$-components of ion displacements (representing shifted equilibrium positions in the presence of $n_{ph}$) onto all $A_1$ modes (see details in SM), it is clear that ions are mostly driven along the eigenvector of the second $A_1$ mode for all $n_{\rm{ph}}$. This mode describes two layer of ions approaching each other (see illustration in the inset of Fig. 2(a)). The projection on the fourth $A_1$ mode, which is about an opposite motion of sulfur atoms within the same monolayer (see illustration in SM), also increases notably when $n_{ph} > 0.1 \ e$/u.c.. Along with the out-of-plane ion motions, there is also an expansion of in-plane constants under illumination, known as the photostriction effect~\cite{kundys2015photostrictive,chen2021photostrictive}, and it is given in SM.
\begin{figure*}
\includegraphics[width = 120mm]{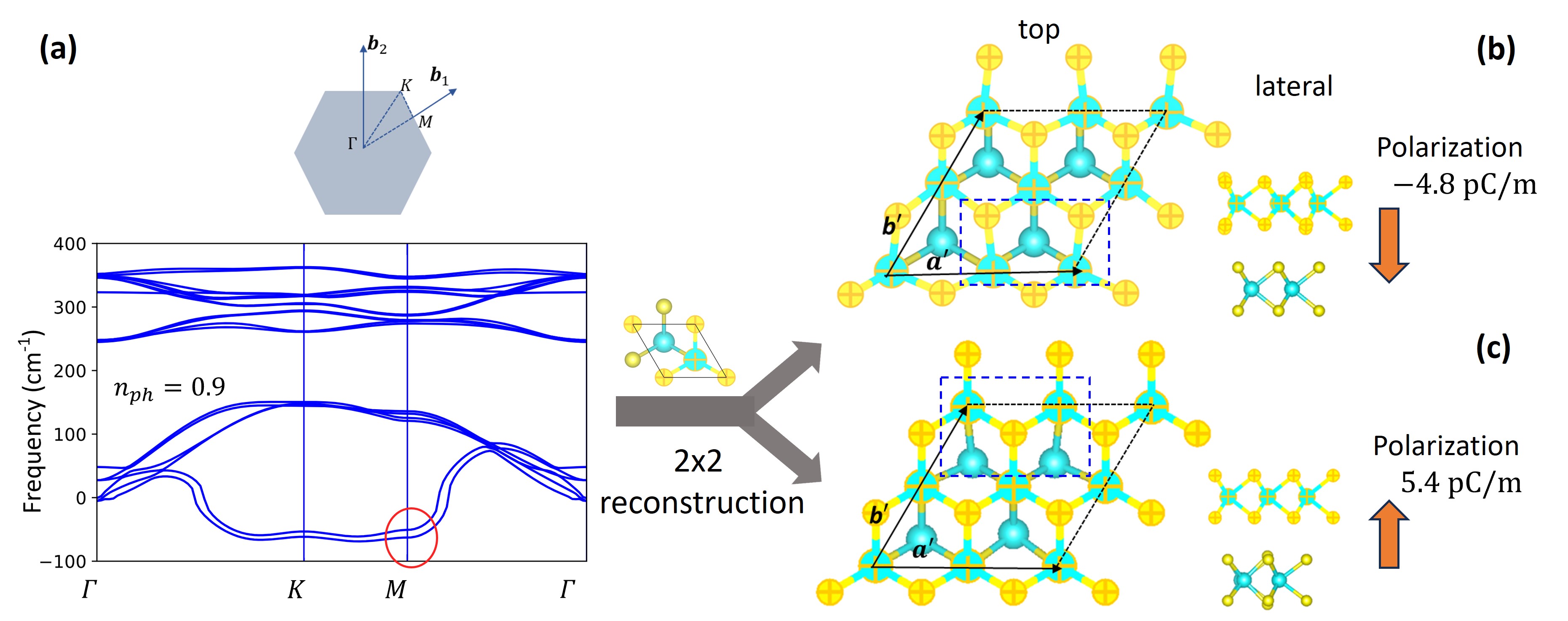}
\caption{(a) Phonon dispersion at $n_{\rm{ph}} =  0.9 \ e /$u.c.. Brillouin zone with high symmetry points is also illustrated. (b)(c) The top and lateral views of two well-relaxed structures based on $2\times2$ reconstruction. Atoms on top layers are highlighted. In (b)/(c), polarization points downwards/upwards, and the structural distortion is in the top/ bottom layer, respectively.}
\end{figure*}
To elucidate the origin of the photo-modulated polarization, we calculate $P_z$ for structures with the same atomic configuration as those well relaxed at all $n_{\rm{ph}}$, but without introducing any e-h pairs. We refer to these $P_z$ as ``structural" contributions, as they solely represent the polarization of relaxed structures under light. Moreover, the difference between these structure-contributed $P_z$ and the overall $P_z$ shown in Fig. 1(b) represent the ``charge" contributions from thermalized e-h pairs created by photoexcitation. The variation of the two types of $P_z$ with $n_{\rm{ph}}$ are shown in Fig. 2(b). When $n_{\rm{ph}}$ varies across $0.01-0.8 \ e /$u.c., the structure-contributed part does not change much and is bound within $0.8-1.05 \ \rm{pC}/\rm{m}$. In contrast, the charge-contributed part varies over a much wider range, increasing from $-0.4 \ \rm{pC}/\rm{m}$ to  $\sim -1 \ \rm{pC}/\rm{m}$  before decreasing to $-0.3 \ \rm{pC}/\rm{m}$. Thus, it plays a more dominant role in tuning $P_z$.

To gain more insight into e-h pairs, we compute the electronic band structure at $n_{\rm{ph}} = 0.2 \  e /$u.c. (see computational details in SM)~\cite{fiorentini1995dielectric,cheiwchanchamnangij2012quasiparticle}. Figure 2(c) shows that photoexcited electrons occupy the bottom part of the two lowest conduction bands near $K$ point. For excited holes, since the lower Fermi level is situated right above the second highest valence band, they reside near the top of the highest valence band around the $\Gamma$ point. This occupation determines a charge distribution that maximizes charge-contributed $P_z$.  Nevertheless, the situation changes when more $n_{\rm{ph}}$ are introduced. In Fig. 2(d), band structure at $n_{\rm{ph}} = 0.8 \  e /$u.c. shows that additional electrons continue to populate the two lowest conduction bands, while additional holes begin to populate the second highest valence band. By occupying these states, associated charge distribution opposes the $P_z$ contributed by the former $0.2 \  e /$u.c. e-h pairs, and it assists the recovery of the overall polarization. This differs markedly from the charge doping in conventional ferroelectrics, where doped carriers consistently suppress the ferroelectric displacements~\cite{kolodiazhnyi2010persistence,wang2012ferroelectric,xia2019coexistence}. For illustration, the occupied orbital characters and charge isosurfaces are given in SM.     

It is well understood that the polarization in sliding ferroelectrics originates from an asymmetric charge transfer between two weakly coupled, inequivalent layers~\cite{li2017binary,yang2018origin,wu2021sliding}. 
We thus follow former analysis and compute the charge density difference ($\rho_{\rm{CDD}}$) to measure the charge transfer~\cite{wang2021vaspkit,rogee2022ferroelectricity} (see computational details in SM).
Figure 2(e) shows $\rho_{\rm{CDD}}(z)$ of bilayer MoS$_{2}$ at dark condition, where charge transfer occurs most in the interlayer region. A small number of transferred charges on top and bottom surfaces also suggests a small depolarization field. This is distinct from photoexcitation in three-dimensional ferroelectrics, where depolarization field from bound charges at polarization domain boundaries should be considered~\cite{abalmasov2020ultrafast,abalmasov2021ferroelectric}. Whereas at $n_{\rm{ph}} = 0.2 \  e /$u.c., Fig. 2(f) shows that transferred charges are most concentrated on intralayer region.
Since the asymmetric charge distribution respective to $z = 0$ plane is responsible for the polarization, we define an asymmetric charge transfer $\rho_{\rm{asym}} (z > 0 ) = \rho_{\rm{CDD}} (z)- \rho_{\rm{CDD}} (-z)$ following Ref.~\citenum{deb2022cumulative}, and we plot $\rho_{\rm{asymm}}(z)$ at different $n_{\rm{ph}}$ (see Figs. S11-S13 in SM).  Indeed, asymmetry in charge distribution is more significant at dark condition, and it is alleviated by e-h pairs at $n_{\rm{ph}} = 0.2 \  e /$u.c.. 

Given the intriguing ferroelectric properties tuned by photodoping, it is important to study whether 3R-stacked bilayer MoS$_2$ can maintain structural stability under light. We thus perform phonon calculations for well relaxed structures at different $n_{\rm{ph}}$~\cite{togo2015first}. No soft modes are observed in phonon dispersions for $n_{\rm{ph}}$ ranging between $ 0.01\sim 0.8 \  e /$u.c. (shown in SM), indicating stability in these structures. However, phonon frequencies at $K $ and $M$ points are gradually reduced with an increasing $n_{\rm{ph}}$, suggesting a potential photo-induced phase transition~\cite{paillard2019photoinduced}. As shown in Fig. 3(a), when $n_{\rm{ph}}$ reaches $0.9 \  e /$u.c., phonon dispersion displays imaginary frequencies at $K$ and $M$ points, demonstrating that the structure becomes unstable. Based on eigenvectors of soft phonon modes at $K$ and $M$ points, we search for the ground state structure in corresponding enlarged supercells, and two well-relaxed, stable structures are derived, as shown in Fig. 3(b)(c) (see computational details in SM). Both structures are based on $2\times 2$ cell reconstruction, and they arise from two soft modes at $M$ point. As highlighted by the dashed boxes, the distortion is only on hexagons of one layer and it involves an alternate shortening and elongation of neighbouring Mo-Mo and S-S distances along $\bm{a'}$ axis; in Fig. 3(b), the distortion is in the top layer and in Fig. 3(c), it is in the bottom layer. In addition, there is also an out-of-plane rippling of the S atoms positioned between two shortened Mo atoms. Such structures have a lower $Cm$ symmetry and cannot be accessed through general stacking or thermalization.  

We compute the polarization of the two $Cm$ structures and find that the top-distorted and the bottom-distorted structures exhibit opposite $P_z$ of $-4.8 \ \rm{pC/m}$ and $5.4 \ \rm{pC/m}$, respectively.
These are almost 5 times greater than the $P_z$ of the original $P3m1$ structure. Similar distortions are also identified in the two ground-state structures at $n_{\rm{ph}}  = 1.0 \  e /$u.c., with $P_z$ of $-4.5 \ \rm{pC/m}$ and $5.3 \ \rm{pC/m}$. The giant enhancement of polarization can be explained by considering symmetry: when one layer experiences distortion as shown in Fig. 3, it becomes more dissimilar to the other layer; this leads to a charge redistribution with a substantial increase in asymmetry along the out-of-plane direction, resulting in a much larger polarization. The asymmetry can also be reversed if the distortion occurs in the other layer. Note that since the $P3m1$ structure inherently carries an upward polarization, we observe two opposite polarizations with nonequivalent values. By decomposing the total polarization into structural and charge contribution, we find that both components are significant (detailed values are given in SM). This suggests that at these critical $n_{\rm{ph}}$, the giant polarization are contributed by both origins.    

In conclusion, using first-principles calculations, we predict a wide range, photo-induced tuning of sliding ferroelectricity in 3R-stacked bilayer MoS$_2$. The  polarization is contributed by a structural component related to structures distorted by light, and a charge component from the distribution of photoexcited carriers; the latter plays a more dominant role in tuning the overall polarization. When light fluence reaches a critical value, a hidden phase is discovered, showing significant distortions and large polarization. Recent experiments suggest that such photo-induced effect remains at least a few tens of picoseconds after the laser pulse~\cite{mannebach2014ultrafast,chernikov2015population,huang2022observation}, and a steady-state, high-density electron-hole plasma can be generated even with continuous wave laser~\cite{wang2019optical}. Optical control of polarization can contribute to the development of nonvolatile, fast-speed data processing and memory devices~\cite{guo2021recent}, while ultrafast structural transition under light presents MoS$_2$ as a promising phase change material to be applied in neuron-inspired
computation~\cite{zhang2019designing}. Our study thus highlight the pivotal role of light, which can be important in engineering the next generation of low-dimensional ferroelectrics.  

{\em Acknowledgement}---We thank Peng Chen, Zhenyao Fang, Yuhao Fu, Lei Li, Charles Paillard, Di Wu and Bin Xu for useful discussions. We are greatly indebted to Giovanni Marini for his generous advice on using constrained DFT implemented in Quantum Espresso. We acknowledge the support from the Grant MURI ETHOS W911NF-21-2-0162 from Army Research Office (ARO) and the Vannevar Bush Faculty Fellowship (VBFF) Grant No. N00014-20-1-2834 from the Department of Defense. We also acknowledge the computational support from HPCMP Pathfinder Award for computational resources and the Arkansas High Performance Computing Center.
 
\bibliography{reference.bib}



\section*{Supplementary material}

# Large photo-induced tuning of ferroelectricity in sliding ferroelectrics

Lingyuan Gao[1] and Laurent Bellaiche[1, *]

[1]*Smart Ferroic Materials Center, Physics Department*
*and Institute for Nanoscience and Engineering,*
*University of Arkansas, Fayetteville, Arkansas, 72701, USA*

(Dated: September 23, 2024)

* laurent@uark.edu

The supplementary material consists of six sections:

(I) Computational details about constrained density functional theory and relevant experimental effects

(II) Benchmark of polarization calculations

(III) Photostriction effect and projection of photo-induced ion displacements onto phonon modes

(IV) Electronic band structure and phonon band structure

(V) Orbital character analysis and charge density difference analysis

(VI) Searching for ground-state structures at high numbers of photoexcited carriers and potential implication in application

# I. COMPUTATIONAL DETAILS ABOUT CONSTRAINED DENSITY FUNCTIONAL THEORY AND RELEVANT EXPERIMENTAL EFFECTS

We use the norm-conserving pseudopotentials with generalized gradient approximation (GGA)[1,2], and the DFT-D2 functional of Grimme is included to correct the vdW interaction[3]. We set up a simulation cell with a dimension of 25 Å along the out-of-plane direction. Accordingly, a $12 \times 12 \times 1$ $\boldsymbol{k}$-point mesh is used for sampling the Brillouin zone[4]. The kinetic cutoff energy for the plane-wave basis is set to 100 Ry, and a smearing parameter of 0.01 Ry is used. To ensure the accuracy, the energy convergence threshold for self-consistent calculations is set as $10^{-11}$ Ry/cell, and the force threshold for structural relaxation is set as $10^{-5}$ Ry/bohr. Spin-orbit coupling effect (SOC) in $MoS_2$ is not included as its strength is moderate compared to other transition metal dichalcogenides and it has a limited effect on electric polarization[5,6]. The translation pathway is computed using the nudged elastic band (NEB) method[7]. As stated in the main manuscript, the number of photoexcited carriers (electron-hole (e-h) pairs) can be estimated from the light fluence $F$ by[8,9]:

$$n_{\rm ph} = \frac{(1-R)\alpha F}{\hbar \omega}, \tag{S1}$$

where $R$, $\alpha$ and $\hbar\omega$ correspond to reflectivity, absorption coeffcients, and photon energy, respectively, and $n_{\rm ph}$ is the volume density (rather than area density) of photoexcited carriers. Here we calculate $\alpha$ and $R$ to have a more consistent prediction. As a benchmark, we first perform calculations on monolayer 2H-$MoS_2$. Following Ref. 10, given the complex

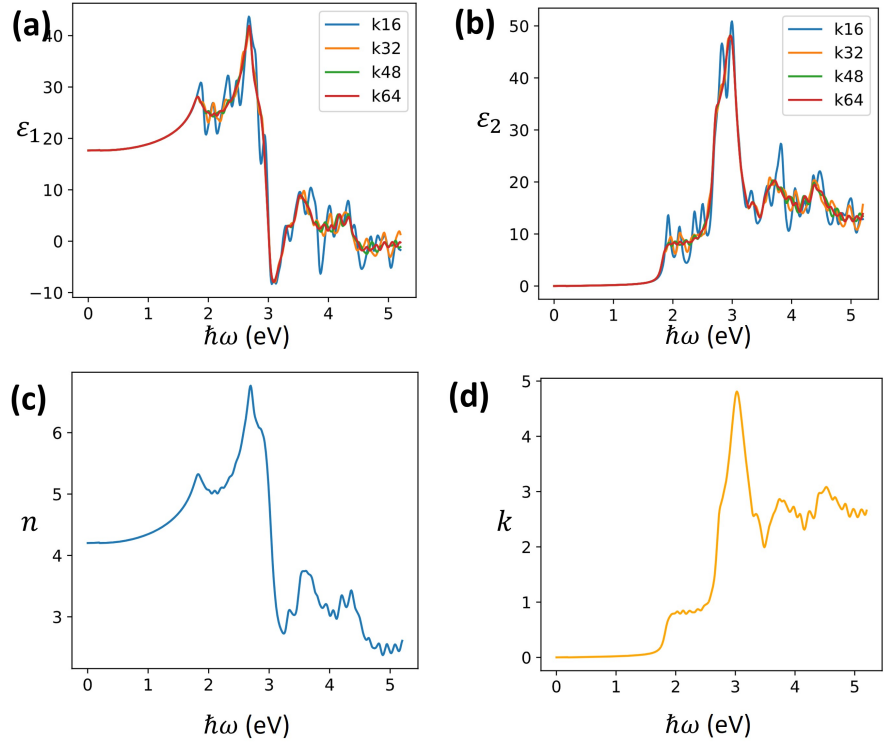

FIG. S1. Complex dielectric constant of monolayer 2H-MoS$_2$ with different Kpoint meshes. (a) the calculated $\varepsilon_1(\omega)$, (b) the calculated $\varepsilon_2(\omega)$, (c) the calculated refractive index $n$, and (d) the calculated extinction coefficient $k$.

dielectric constant $\varepsilon_1(\omega)$ (real part) and $\varepsilon_2(\omega)$ (imaginary part), the refractive index $n$ and extinction coefficient $k$ are expressed as:

$$n(\omega) = \sqrt{\frac{1}{2}(\sqrt{\varepsilon_1(\omega)^2 + \varepsilon_2(\omega)^2} + \varepsilon_1(\omega))}, k(\omega) = \sqrt{\frac{1}{2}(\sqrt{\varepsilon_1(\omega)^2 + \varepsilon_2(\omega)^2} - \varepsilon_1(\omega))}, \quad (S2)$$

and the reflectivity $R(\omega)$ and absorption coefficient $\alpha(\omega)$ are given as:

$$R(\omega) = \frac{(1-n(\omega))^2 + k(\omega)^2}{(1+n(\omega))^2 + k(\omega)^2}. \quad (S3)$$

Below we computed $\varepsilon_1(\omega)$ (real part) and $\varepsilon_2(\omega)$ first. The calculation of $\varepsilon(\omega)$ requires a fine Kpoint mesh as wavefunctions are highly involved in computing momentum matrix elements[11]. In addition, scissor operation[12], as introduced in section IV, is also applied to correct the band gap. As shown in Fig. S1(a)(b), the calculated $\varepsilon_1(\omega)$ and $\varepsilon_2(\omega)$ are well converged with a 64×64×1 Kpoint mesh. Qualitatively, they agree well with experimentally measured dielectric constants[13], and the discrepancy results from the independent particle approximation used here compared to quasiparticle self-consistent $GW$ method[14,15]. We

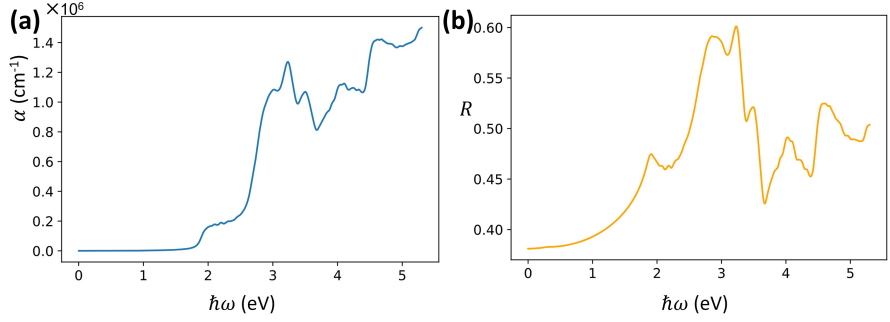

FIG. S2. (a) Calculated absorption coefficient $\alpha(\omega)$ (cm$^{-1}$), and (b) calculated reflectivity $R$ of bilayer 3R-MoS$_2$.

also present calculated $n$ and $k$ in Fig. S1(c)(d). Close match between calculation and experimental results given in Ref. 13 indicates that it is safe to apply the presently used computational method on the bilayer 3R-MoS$_2$.

The calculated absorption coefficient $\alpha(\omega)$ and reflectivity $R(\omega)$ are presented in Fig. S2. If we choose the energy of pump photon $\hbar\omega$ at 3 eV, so that the photoexcitation is well above the band edge and does not resonate with the exciton, the corresponding $\alpha(\omega)$ and $R(\omega)$ are $1.1 \times 10^6$ cm$^{-1}$ and 0.57. Note that the calculated $\alpha$ for bilayer 3R-MoS$_2$ is at the same order as $\alpha$ measured for monolayer 2H-MoS$_2$[16–18]. With these values, two critical fluences $F$ at $n_{\rm ph} = 0.2$ $e$/u.c. (equivalent to $n_{\rm 2D} \approx 2 \times 10^{14}$/cm$^2$) and $n_{\rm ph} = 0.9$ $e$/u.c. ($n_{\rm 2D} \approx 10^{15}$/cm$^2$) are estimated to be 1.8 mJ/cm$^2$ and 8.2 mJ/cm$^2$, respectively.

We note that the effect of photoexcitation on structural instability is very different from static heating, with the latter being directly related to the phase transition temperature. Based on Refs. 19,20, the 2H-1T phase transformation in monolayer MoS$_2$ occurs at T = 600 °$C$, and the associated structural transition under heat is not reversible. Several experiments report that the monolayer MoS$_2$ exhibits structural stability under fluences at the order of mJ/cm$^2$: Reference 21 reported that with an incident photon energy of 1.55 eV, the monolayer MoS$_2$ remains unaffected by a single pulse of 50 mJ/cm$^2$ for and by a train of 10$^3$ pulses of 20 mJ/cm$^2$, as revealed by optical microscopy image and field emission scanning electron microscopy. Reference 22 reported that when monolayer MoS$_2$ is exposed to an incident photon energy of 3.6 eV at a fluence of 10 mJ/cm$^2$, large-amplitude increases in the time-resolved second harmonic response emerge following the pump pulse, and the signal recovers at tens of picoseconds (ps) level; this highly contrasts to the irreversible

transformation at T = 900 °C by static heating[22]. Moreover, Reference 23 reported that the atomic force microscopy image and Raman spectrum of monolayer MoS$_2$ flake remain the same after the flake being exposed to 1.85 eV irradiation at a fluence of 50 mJ/cm$^2$ for 10 minutes. All these studies demonstrate that the heat effect associated with above-bandgap-photoexcitation at a fluence 10 mJ/cm$^2$ on the structure of monolayer MoS$_2$ is very limited. Moreover, creating dense e-h plasma with intense above-bandgap-photoexcitation has been realized in transition metal dichalcogenides system: it is estimated that $10^{15}$/cm$^2$ e-h pairs are created by the 10 mJ/cm$^2$ pump photoexcitation in Ref. 22. Reference [13] also showed that a n$_{2D}$ = 1.1 × $10^{14}$/cm$^2$ e-h plasma is created in monolayer and bilayer WS$_2$ with a 2.4 eV pump laser pulse at a fluence of 0.87 mJ/cm$^2$ (which is close to our estimate about the fluence vs n$_{ph}$), and a higher density is achieved as the fluence increases up to 3.4 mJ/cm$^2$; later the e-h plasma recovers to the initial phase after a few hundred ps. Furthermore, Reference 24 reported that e-h plasma with carrier density up to 4 × $10^{14}$/cm$^2$ is created in hetero-bilayer WSe$_2$/MoSe$_2$ system with both pulsed and continuous wave photoexcitation at 2.3 eV. These experiments demonstrate that our theoretical proposal to tune ferroelectricity and drive structural phase transition with photoexcited carriers at ∼$10^{14}$/cm$^2$ in bilayer 3R-MoS$_2$ is practical and experimentally achievable.

## II. BENCHMARK OF POLARIZATION CALCULATIONS

Nowadays electric polarization is typically computed using the Berry phase method[25–27]. Nevertheless, the Berry phase (BP) method implemented in Quantum Espresso requires a full occupation on valence bands[28–30], which is not satisfied in the photodoped case. Therefore, we seek an alternative approach to compute the polarization $\vec{P}$ as the integration of charge density $n(\boldsymbol{r})$:

$$\vec{P} = \frac{1}{V} \int n(\boldsymbol{r})\boldsymbol{r}d\boldsymbol{r}, \tag{S4}$$

where $V$ is the volume of the system. Although this method cannot compute polarization for a periodic system due to the ambiguity of polarization with different origin choices, it is well-defined for non-periodic systems, such as the out-of-plane direction of a slab[31].

Another important point for slab calculation is the dipole correction, which has been used to remove the artificial electric field in the vacuum region of the computational cell[32,33]. Nevertheless, this correction affects the charge distribution as well as the polarization of

5

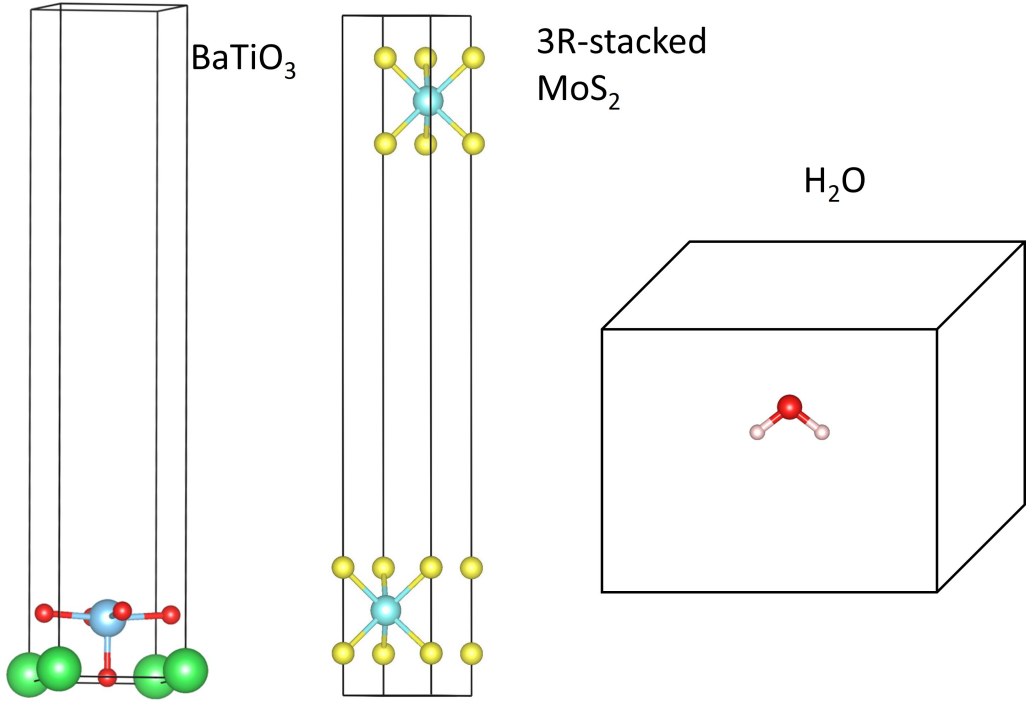

FIG. S3. Structures of a one-layer well-relaxed BaTiO$_3$ slab, 3R-stacked bilayer MoS$_2$, and a water molecule H$_2$O for benchmarking polarization calculations.

|  | BP method without DC ($e \cdot$ Å/u.c.) | CCI method without DC ($e \cdot$ Å/u.c.) | BP method with DC ($e \cdot$ Å/u.c.) | CCI method with DC ($e \cdot$ Å/u.c.) |
| --- | --- | --- | --- | --- |
| BaTiO$_3$ | 0.3350 | 0.3349 | 0.2756 | 0.2757 |
| 3R-MoS$_2$ | 0.0045 | 0.0046 | 0.0027 | 0.0027 |
| H$_2$O | 0.3803 | 0.3802 | 0.3784 | 0.3783 |

TABLE I. Electric dipole moments (in the unit of $e \cdot$ Å/unit cell (u.c.)) of one-layer BaTiO$_3$ slab, 3R-stacked bilayer MoS$_2$, and H$_2$O molecule calculated by four different methods: BP method without DC; CCI method without DC; BP method with DC; CCI method with DC.

the slab. Below, we examine the classical charge integration (CCI) method in polarization calculation, and the dipole correction (DC) effect on polarization, in three different types of systems: a well-relaxed one-layer BaTiO$_3$ slab, 3R-stacked bilayer MoS$_2$, and a water molecule H$_2$O. All three systems are polar and exhibit electric dipole moments along the out-of-plane direction, and their structures are illustrated in Fig. S3.

Table I presents the computed electric dipole moments $\vec{p} = V\vec{P}$ for all three systems

6

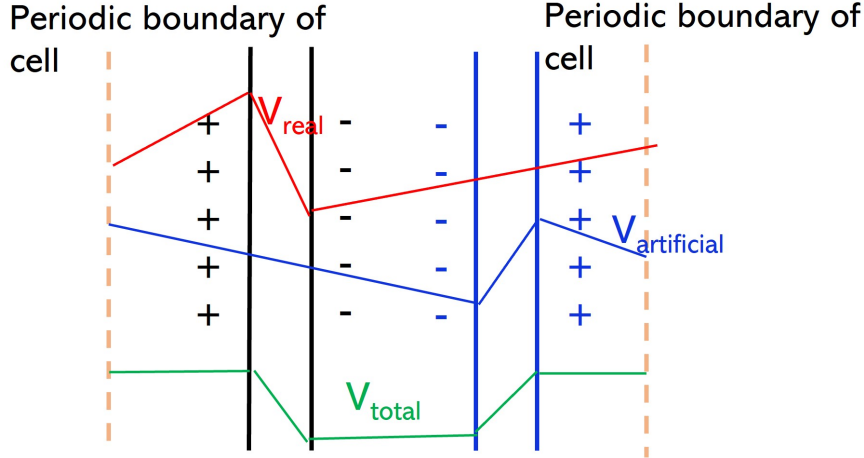

FIG. S4. Illustration about different electrostatic potentials under dipole correction. The red line ($V_{\text{real}}$) denotes the electrostatic potential of the original slab (denoted by two vertical black lines); the blue line ($V_{\text{artificial}}$) denotes the added sawtooth potential, equivalently contributed by the artificial slab (denoted by two vertical blue lines); the green line (($V_{\text{total}}$)) denotes the overall electrostatic potential as a sum of $V_{\text{real}}$ and $V_{\text{artificial}}$.

using four different methods: BP method without DC, CCI method without DC, BP method with DC, CCI method with DC. Both results with and without DC given by BP and CCI methods agree very well, demonstrating that the polarization computed by CCI method is reliable. Nevertheless, we see a notable difference between results with and without DC: when DC is added, $\vec{p}$ decreases by 20% for BaTiO$_3$ slab and by 40% for 3R MoS$_2$, while $\vec{p}$ of H$_2$O decreases by less than 1%.

The DC effect on polarization can be understood as follows. As illustrated in Fig. S4, there is a drop of the electrostatic potential $V_{\text{real}}$ within the slab owing to its polarity. To satisfy the periodic boundary condition and to have a continuous $V_{\text{real}}$ throughout the cell, $V_{\text{real}}$ in vacuum regions varies to reach the same value at the boundary of the cell from both sides of the slab. The variation of $V_{\text{real}}$ corresponds to an electric field, which does not exist in the real circumstance. To cancel this field in vacuum regions, an artificial sawtooth potential $V_{\text{artificial}}$ is added, and it leads to an overall electrostatic potential $V_{\text{total}}$ with a plateau in vacuum regions. This is equivalent to placing an artificial slab with opposite polarity far from the original slab, and the electric field introduced by the artificial slab opposes the electric dipole moment of the original slab, giving rise to smaller $\vec{p}$ values when

7

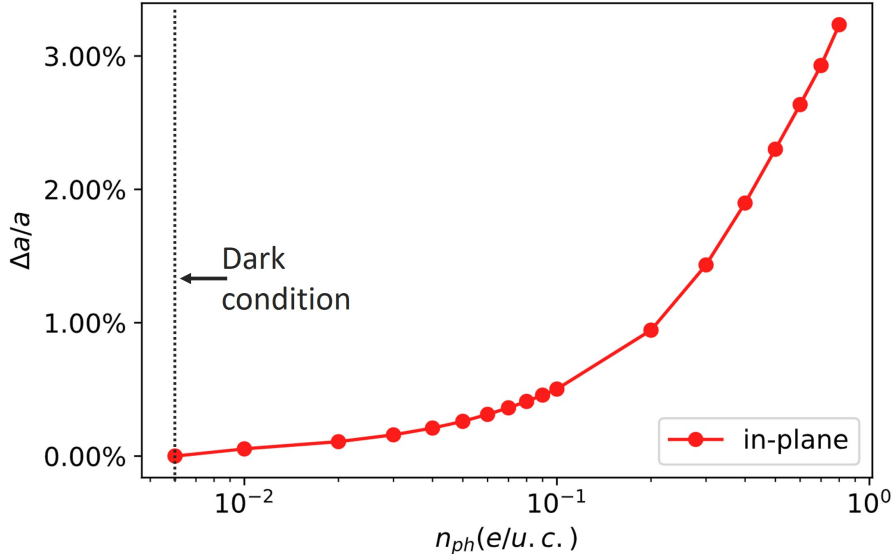

FIG. S5. Photostriction effect showing the lattice expansion of in-plane lattice constant with the increasing photoexcited carriers $n_{\text{ph}}$.

dipole correction is turned on. We see that the effect is much stronger in periodic two-dimensional systems than in an isolated molecule; this is because the electric field of an infinite slab (a constant proportional to the area charge density) is much stronger than that of a point charge.

In the main manuscript, we report polarization values calculated without DC; in this way, the electric polarization is not affected by an artificial electric field. We would like to emphasize that polarizations should be compared consistently, i.e., either all with DC or all without DC.

## III. PHOTOSTRICTION EFFECT AND PROJECTION OF PHOTO-INDUCED ION DISPLACEMENTS ONTO PHONON MODES

At dark condition, the in-plane lattice constant $a$ of 3R-stacked bilayer $MoS_2$ is 3.214 Å. With the increasing photoexcited carriers $n_{\text{ph}}$, there is a lattice expansion of in-plane lattice constant. As shown in Fig. S5, the increase is about 3.2% when $n_{\text{ph}} = 0.8$ $e$/u.c., comparable to photostriction in hybrid improper ferroelectric superlattice[34]. For $n_{\text{ph}} > 0.8$ $e$/u.c., $2 \times 2$ lattice reconstructions occur, and lattice constants of reconstructed cells are $a = 6.543$ Å and 6.556 Å for $n_{\text{ph}} = 0.9$ and 1.0 $e$/u.c. respectively.

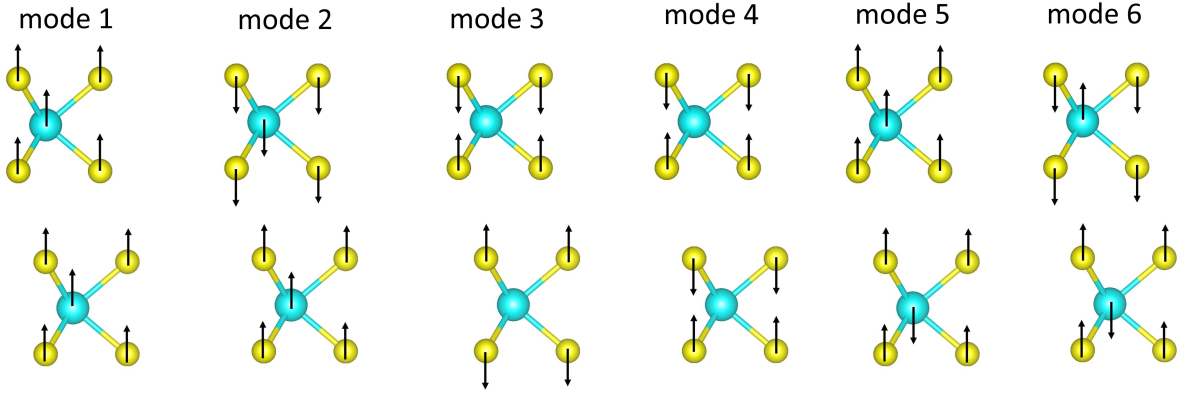

FIG. S6. Illustration about 6 all-symmetric $A_1$ phonon modes at $\Gamma$ point ($\bm{q} = 0$) from the lateral view. Mo and S atoms are denoted by blue balls and yellow balls, respectively.

To characterize the "displacive excitation of coherent phonons" (DECP) mechanism[35–41], we first extract eigenvectors of 6 $A_1$ phonon modes at $\Gamma$ point and illustrate their motions in Fig. S6 in the order of an increasing frequency, which are all along the out-of-plane $z$ direction. We denote the equilibrium position of atom $j$ at dark condition as $(x_0^j, y_0^j, z_0^j)$, and its equilibrium position at $n_{\mathrm{ph}}$ as $(x_{n_{\mathrm{ph}}}^j, y_{n_{\mathrm{ph}}}^j, z_{n_{\mathrm{ph}}}^j)$, so the out-of-plane ion displacement $D_z^j(n_{\mathrm{ph}})$ driven by photoexcitation is $z_{n_{\mathrm{ph}}}^j - z_0^j$. The six $A_1$ phonon modes form a complete atomic basis for out-of-plane motions of 6 atoms in this system. In addition, the eigenvector $\bm{e}_\lambda$ of phonon mode $\lambda$ at $\Gamma$ point ($\bm{q} = 0$) corresponds to a real-space displacement $\bm{u}_\lambda^j \propto \frac{1}{\sqrt{m_j}} \bm{e}_\lambda^j$ at atom $j$, and $m_j$ is the atomic mass. Therefore, we renormalize the ion displacement $D_z^j(n_{\mathrm{ph}})$ by multiplying a mass factor $\sqrt{m_j}$, and decompose it as the summation of all $A_1$ modes:

$$\sqrt{m_j} D_z^j(n_{\mathrm{ph}}) = \sum_{\lambda=1}^{6} C_{\mathrm{proj}}^\lambda(n_{\mathrm{ph}}) \bm{e}_\lambda^j, \tag{S5}$$

where $j$ runs from 1 to 6, and $C_{\mathrm{proj}}$ denotes the projection coefficient on mode $\lambda$. Such projection naturally yields $\sum_{\lambda=1}^{6} |C_{\mathrm{proj}}^\lambda(n_{\mathrm{ph}})|^2 = 1$. As shown in Fig. 2(a) of the main manuscript, the large $|C_{\mathrm{proj}}^{\lambda=2}|$ indicates that the out-of-plane ion displacements under light are mostly driven along the eigenvector of the second $A_1$ mode throughout all $n_{\mathrm{ph}}$. When $n_{\mathrm{ph}} > 0.1$ $e$/u.c., the increasing $|C_{\mathrm{proj}}^{\lambda=4}|$ indicates that ion displacements are also partially driven along the eigenvector of the fourth $A_1$ mode. The fourth $A_1$ mode describes an opposite motion of sulfur atoms within the same monolayer, and sulfur atoms in the interlayer region move away from each other.

9

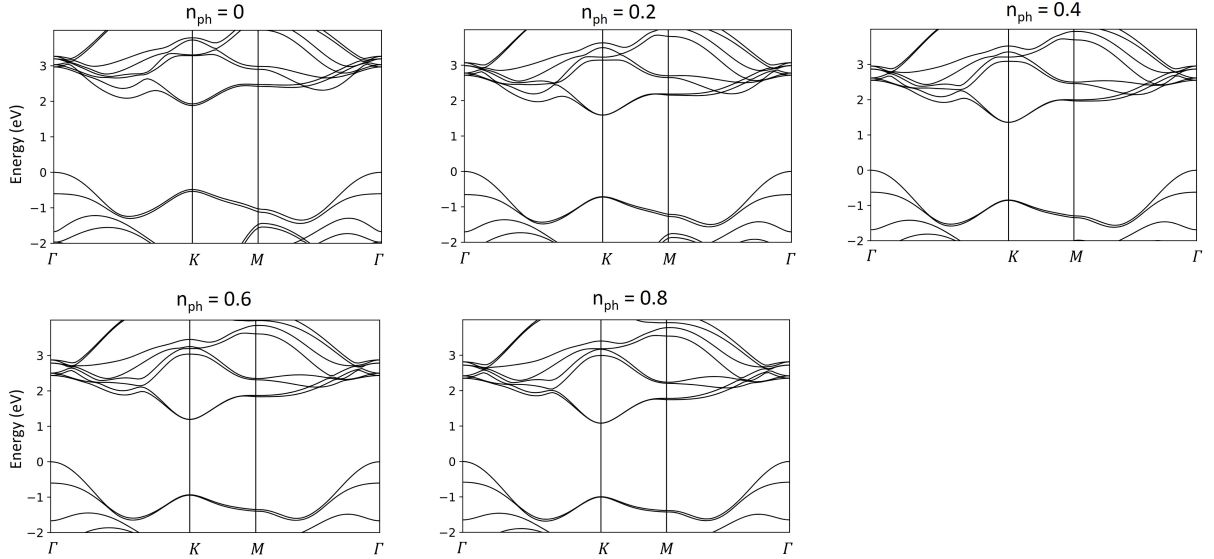

FIG. S7. Scissor-corrected electronic band structures at dark condition and at different $n_{\text{ph}}$.

## IV. ELECTRONIC BAND STRUCTURE AND PHONON BAND STRUCTURE

As explained in the main manuscript, when the photocarrier density is larger than $10^{13}$ cm$^{-2}$ ($\sim 0.01$ $e$/u.c., ), electron-hole plasma forms, and Coulomb interaction is largely reduced. Therefore, electronic band structures computed under single particle approximation remain valid. However, it is well known that density functional theory can underestimate the band gap of semiconductors due to the self-interaction error[42–44]. To correct the band gap, we first calculate the band structure without photodoping (dark condition) at GGA level; by comparing the band gap with that from quasiparticle self-consistent $GW$ method[14,15], we get a gap difference between two methods. We then apply a scissor operation, i.e., rigidly shifting the conduction bands by this difference[12]. All corrected band structures at different levels of $n_{\text{ph}}$ are shown in Fig. S7. The gap is reduced from 1.88 eV to 1.08 eV from dark condition to $n_{\text{ph}} = 0.8$ $e$/u.c..

For phonon calculation, we construct $4 \times 4 \times 1$ supercells based on well-relaxed, 6-atom unit cells of 3R-stacked bilayer MoS$_2$ at different $n_{\text{ph}}$; $n_{\text{ph}}$ in each supercell is also increased in ratio. We perform self-consistent calculations on supercell configurations generated by the Phonopy software[45], and we extract atomic forces and compute phonon band structure within the frozen phonon method. All phonon band structures at different levels of $n_{\text{ph}}$ are shown in Fig. S8. With an increasing $n_{\text{ph}}$, phonon frequencies are shifted downwards, and

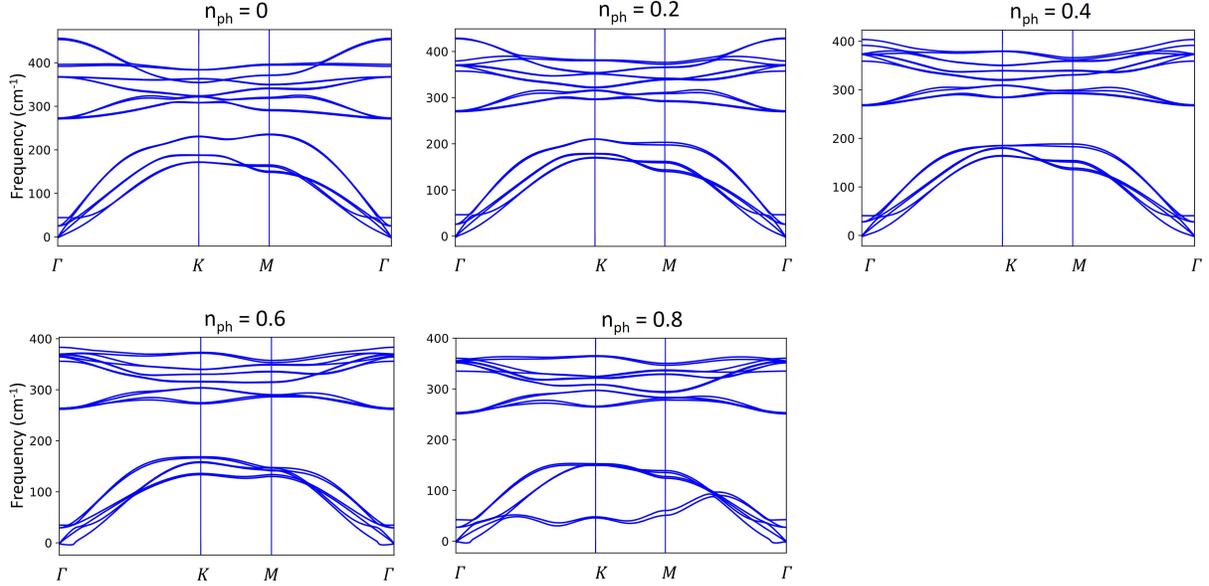

FIG. S8. Phonon band structures at dark condition and at different $n_{\mathrm{ph}}$.

the two lowest branches exhibit strong softening effect at $n_{\mathrm{ph}} = 0.8$ $e$/u.c..

## V. ORBITAL CHARACTER ANALYSIS AND CHARGE DENSITY DIFFERENCE ANALYSIS

We plot the projected density of states (PDOS) at $n_{\mathrm{ph}} = 0.2$ $e$/u.c.. in Fig. S9(b), along with the band structure shown in Fig. S9(a). PDOS reveals that the top of the valence band region is primarily Mo-$d_{z^2}$ and S-$p_z$ orbitals, and the bottom of the conduction band region is primarily Mo-$d_{z^2}$ orbitals; the former and the latter are occupied thermalized holes and electrons, respectively. Respective charge isosurfaces are also given in Fig. S9(a) for demonstrating associated orbital characters. When $n_{\mathrm{ph}} = 0.8$ $e$/u.c.., Fig. S10 reveals that additional electrons further occupy Mo-$d_{x^2-y^2} + d_{xy}$ and S-$p_x + p_y$ states in addition to Mo-$d_{z^2}$ states, and additional holes further occupy Mo-$d_{x^2-y^2} + d_{xy}$ and S-$p_x + p_y$ states alongside Mo-$d_{z^2}$ and S-$p_z$ states.

The charge density difference $\rho_{\mathrm{CDD}}$ can track the asymmetric charge transfer from the monolayer MoS$_2$ component to 3R-stacked bilayer MoS$_2$, which contributes to the resultant out-of-plane polarization. We thus study how $\rho_{\mathrm{CDD}}$ is affected by photoexcitation. As introduced in Refs. 46,47, $\rho_{\mathrm{CDD}}$ is defined as $\rho_{\mathrm{CDD}} = \rho_{\mathrm{BL}} - \rho_{\mathrm{mono-up}} - \rho_{\mathrm{mono-dw}}$, where "BL"

11

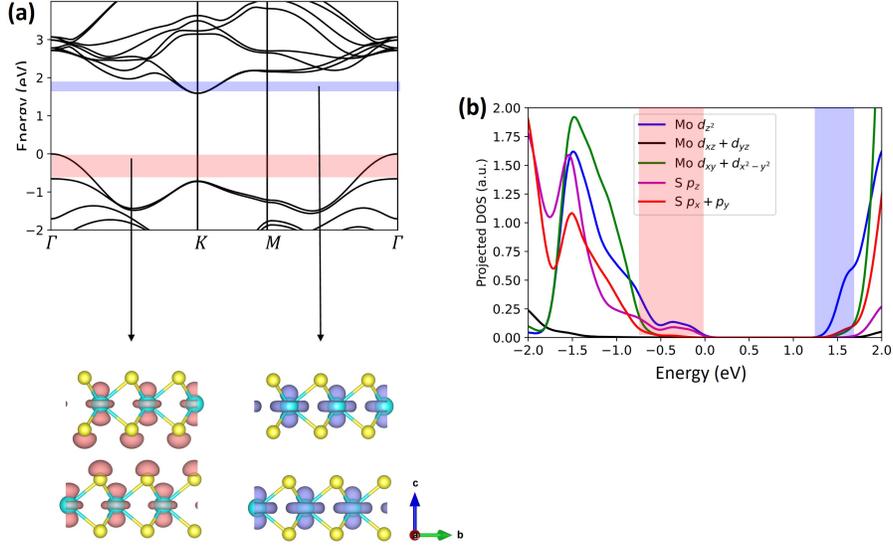

FIG. S9. (a) Electronic band structure at $n_{\mathrm{ph}} = 0.2$ $e$/u.c., with regions occupied by thermalized electrons (blue) and holes (red) highlighted. Charge isosurfaces of these occupied states are also given, with red denoting hole states and blue denoting electron states; the level of hole and electron isosurfaces are at $2 \times 10^{-3}$ $e/V_{\mathrm{grid}}$ and $3 \times 10^{-3}$ $e/V_{\mathrm{grid}}$, where $V_{\mathrm{grid}}$ denotes the volume of a mesh grid ($\approx 3 \times 10^{-3}$ bohr$^3$). (b) Projected density of states at $n_{\mathrm{ph}} = 0.2$ $e$/u.c. Consistent with (a), regions occupied by thermalized electrons and holes are also highlighted.

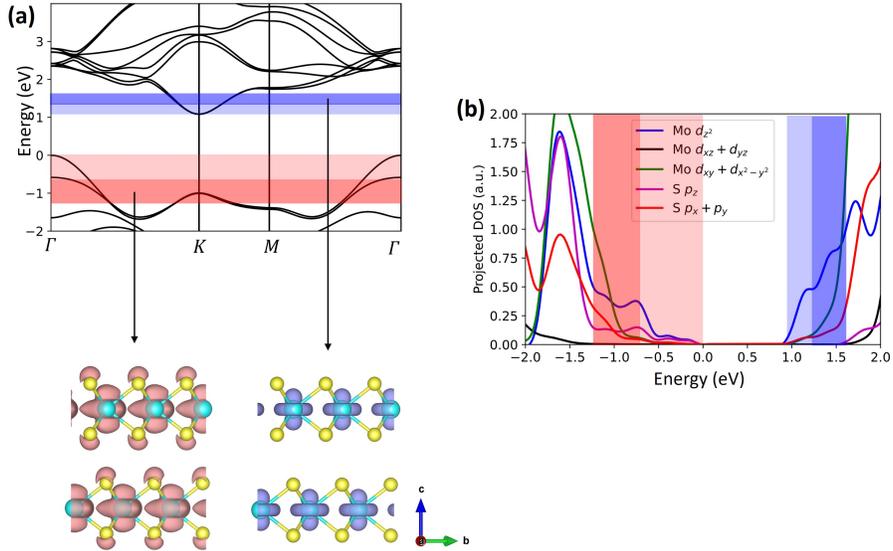

FIG. S10. Similar to Fig. S9, but at $n_{\mathrm{ph}} = 0.8$ $e$/u.c.. Additional thermalized electrons and holes beyond 0.2 $e$/u.c. are marked by darker blue and darker red, and associated charge isosurfaces at the level of $2 \times 10^{-3}$ $e/V_{\mathrm{grid}}$ are also given.

12

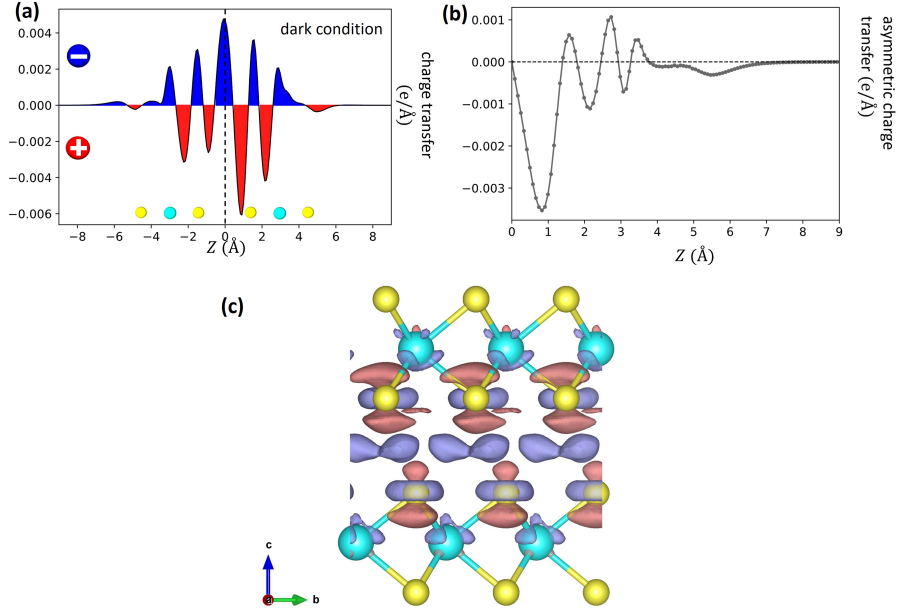

FIG. S11. (a) Charge density difference $\rho_{\text{CDD}}$ at dark condition, as a measure of charge transfer. Blue denotes the electron accumulation and red denotes the electron depletion. (b) Asymmetric charge transfer $\rho_{\text{asymm}}$ at dark condition. (c) Isosurface of $\rho_{\text{CDD}}$ at the level of $1.1 \times 10^{-4}$ $e/V_{\text{grid}}$; purple denotes the electron accumulation and the red brown denotes the electron depletion.

denotes the 3R-stacked bilayer, and "mono-up/dw" denote the independent, up or down monolayer constituents, and $\rho$ is the plane-averaged charge density, contributed by both an ionic part $\rho_{\text{ion}}(z)$ and electronic part $\rho_{\text{elec}}(z)$. We plot $\rho_{\text{CDD}}$ at dark condition along the $\boldsymbol{z}$ axis in Fig. S11(a).

Without writing all constants explicitly, the out-of-plane polarization component $P_z$ can be written as $P_z = \int \rho(z) z dz = \int_{z>0} (\rho(z) - \rho(-z)) z dz = \int_{z>0} \rho_{\text{asymm}}(z) z dz$, and $\rho_{\text{asymm}}(z)$ is defined as the asymmetric charge density between $+z$ and $-z$. Since $\rho_{\text{ion}}(z)$ is symmetric respect to $z = 0$ plane, $\rho_{\text{asymm}}(z)$ is fully determined by $\rho_{\text{elec}}(z)$. Alternatively, we can write $\rho_{\text{asymm}}(z) = \rho_{\text{CDD}}(z) - (\rho_{\text{CDD}})(-z)$. If $\rho_{\text{asymm}}(z)$ maintains the same sign across $z$, its accumulation results in a larger $P_z$ value; otherwise, positive and negative $P_z$ will cancel out during integration, resulting in a smaller $P_z$. We plot $\rho_{\text{asymm}}(z)$ at dark condition along $z$ axis in Fig. S11(b). We illustrate the isosurface of transferred charges in real space in Fig. S11(c).

Figures S12, S13 show $\rho_{\text{CDD}}$ at $n_{\text{ph}} = 0.2$ $e/$u.c. and $0.8$ $e/$u.c., respectively. When

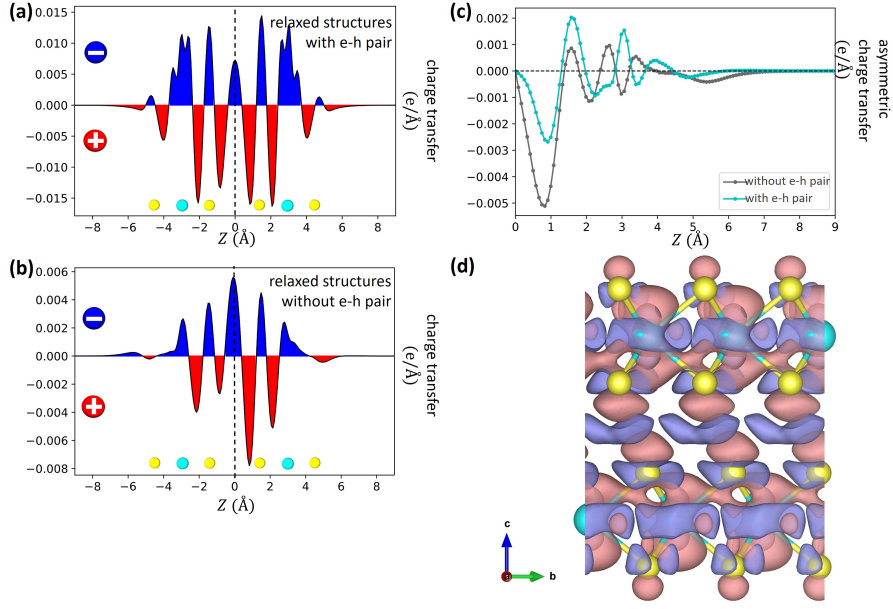

FIG. S12. (a) $\rho_{\text{CDD}}$ corresponding to total polarization at $n_{\text{ph}} = 0.2$ $e$/u.c., and $\rho_{\text{BL}}$ is the charge density of the bilayer structures well relaxed at $n_{\text{ph}}$ (e-h pairs included). (b) $\rho_{\text{CDD}}$ corresponding to structure-contributed polarization at $n_{\text{ph}} = 0.2$ $e$/u.c., and $\rho_{\text{BL}}$ is the charge density of structures with the same bilayer atomic configuration but without any e-h pairs. (c) Asymmetric charge transfer $\rho_{\text{asymm}}$ at $n_{\text{ph}} = 0.2$ $e$/u.c. with and without e-ph pairs. (d) Isosurface of $\rho_{\text{CDD}}$ in (a) at the level of $1.2 \times 10^{-4}/V_{\text{grid}}$.

$n_{\text{ph}} \neq 0$, there are two $\rho_{\text{CDD}}$ needed to be considered: one corresponds to the structure-contributed polarization (Fig. 2(b) in the main manuscript), and the other one corresponds to the total polarization (Fig. 1(b) in the main manuscript); their difference is determined by distribution of the photoexcited carriers. We thus adopt $\rho_{\text{BL}}$ as the charge density of the bilayer structures well relaxed at $n_{\text{ph}}$ for $\rho_{\text{CDD}}$ related to total polarization (e-h pairs included) in Figs. S12(a) and S13(a), and we adopt $\rho_{\text{BL}}$ as the charge density of structures with the same bilayer atomic configuration but without any e-h pairs for $\rho_{\text{CDD}}$ related to structure-contributed polarization in Figs. S12(b) and S13(b). For both $\rho_{\text{CDD}}$, $\rho_{\text{mono}-\text{up/dw}}$ is adopted as the charge density of the respective monolayer component without e-h pairs.

By comparing (a) and (b) of these two figures, it is clear that more charges are transferred in the presence of e-h pairs; nevertheless, as revealed by $\rho_{\text{asymm}}$ in (c) of Figs. S12 and S13, the asymmetry in charge distribution is reduced by e-h pairs, resulting in a smaller overall polarization compared to its structure-contributed components.

14

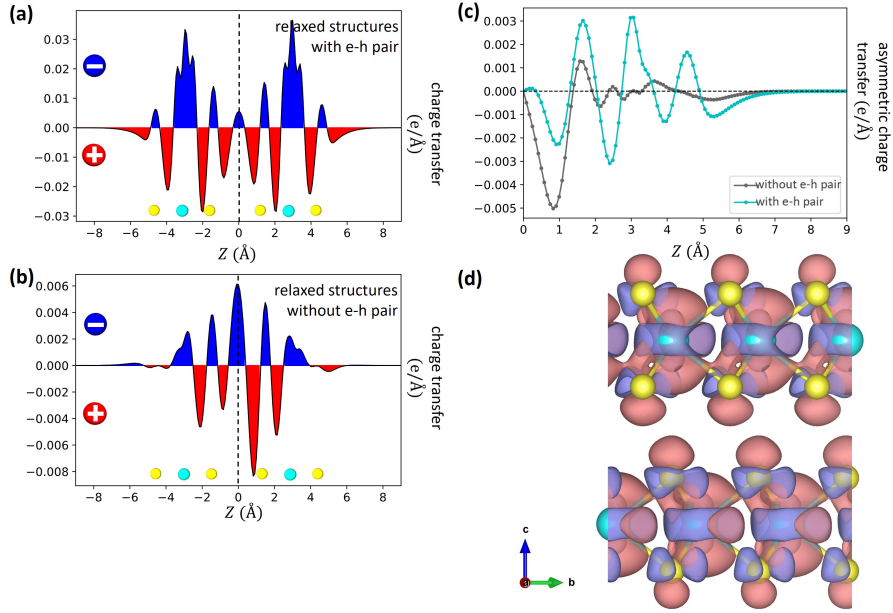

FIG. S13. (a)–(c) Similar to Fig. S12, but at $n_\text{ph} = 0.8$ $e$/u.c.. (d) Isosurface of $\rho_\text{CDD}$ in (a) at the level of $3.7 \times 10^{-4}/V_\text{grid}$.

## VI. SEARCHING FOR GROUND-STATE STRUCTURES AT HIGH NUMBERS OF PHOTOEXCITED CARRIERS AND POTENTIAL IMPLICATION IN APPLICATION

As shown in Fig. 3 of the main manuscript, two lowest phonon mode frequencies at $K$ and $M$ points become imaginary when $n_\text{ph} = 0.9$ $e$/u.c., indicating structural instabilities at this level of $n_\text{ph}$. Figure S14 illustrates two different cell reconstructions induced by soft phonon modes at these high symmetry points: $\sqrt{3} \times \sqrt{3}$ reconstruction corresponds to modes at $K$ point, and $2 \times 2$ reconstruction corresponds to modes at $M$ point. To demonstrate this correspondence, we compute phonon frequencies at $\Gamma, K, M$ points of the primitive cell at $n_\text{ph} = 0.1$ $e$/u.c., and we compare them with phonon frequencies at $\Gamma$ point of the larger reconstructed cells. The computed phonon frequencies $\omega$ are presented in Table II, and we find a good match between $\Gamma$-point frequencies of $2 \times 2$ ($\sqrt{3} \times \sqrt{3}$) cell with $\Gamma$-point and $M(K)$-point frequencies of the primitive cell.

To search for the ground state structure, we first solve eigenvectors of two soft phonon modes at $K$ and $M$ points. Using the modulation function of the Phonopy software[45], these eigenvectors can then be added as initial displacements to atom $j$ in specific enlarged

15

| $\Gamma - \omega$ of $2 \times 2$ (cm$^{-1}$) | $\Gamma - \omega$ of primitive | $M - \omega$ of primitive | $\Gamma - \omega$ of $\sqrt{3} \times \sqrt{3}$ | $K - \omega$ of primitive |
|---|---|---|---|---|
| 29 | 29 | | 31 | |
| 43 | 42 | | 44 | |
| 144 | | 144 | 171 | 171 |
| 148 | | 148 | 172 | 172 |
| 160 | | 160 | 186 | 184 |
| 163 | | 163 | 186 | 184 |
| 214 | | 214 | 221 | 223 |
| 219 | | 219 | 221 | 224 |
| 271 | 271 | | 271 | |
| 273 | 273 | | 273 | |
| 291 | | 291 | 302 | 303 |
| 293 | | 293 | 302 | 304 |
| 314 | | 314 | 319 | 320 |
| 316 | | 316 | 321 | 321 |
| 342 | | 342 | 322 | 323 |
| 343 | | 343 | 324 | 324 |
| 345 | | 345 | 355 | 354 |
| 368 | | 368 | 356 | 355 |
| 369 | | 369 | 361 | 360 |
| | | | 362 | 360 |
| 369 | 369 | | 370 | |
| 370 | 370 | | 371 | |
| 372 | 372 | | 374 | |
| 381 | | 381 | 382 | 383 |
| 384 | | 384 | 383 | 383 |
| 388 | 388 | | 387 | |
| 442 | 442 | | 442 | |
| 445 | 445 | | 445 | |

TABLE II. phonon frequencies at different wavevector $\boldsymbol{q}$ in different cells; acoustic modes are not included, and frequencies of degenerate modes are presented only once.

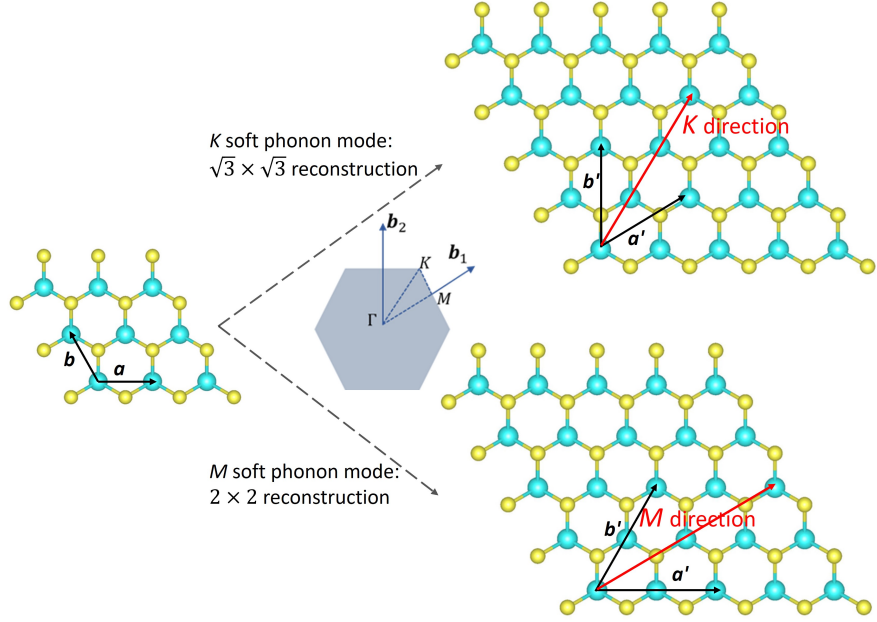

FIG. S14. Illustration about two different cell reconstructions induced by soft phonon modes at $K$ and $M$ points. Only the in-plane view of the top monolayer is presented for clarity. $\boldsymbol{a}, \boldsymbol{b}$ denote the initial in-plane lattice vectors and $\boldsymbol{a'}, \boldsymbol{b'}$ denote the reconstructed in-plane lattice vectors.

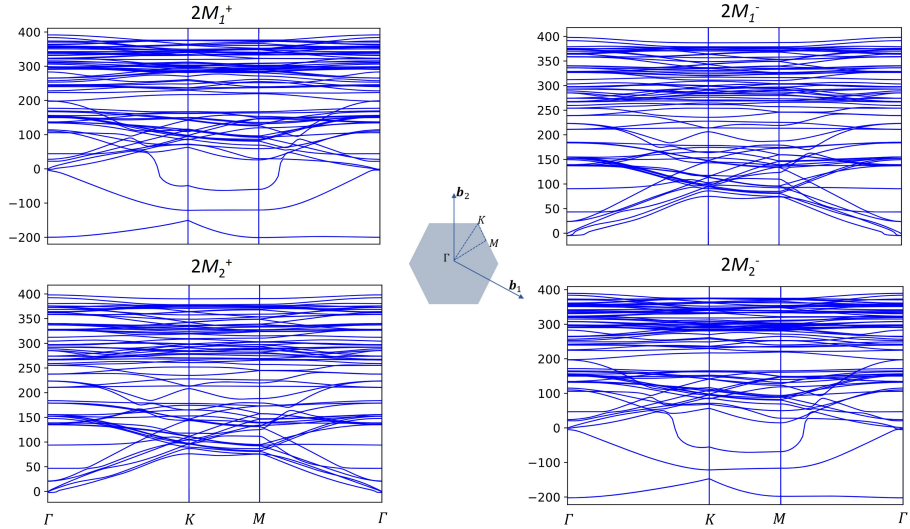

FIG. S15. Phonon band structures of well relaxed $2 \times 2$ reconstructed cells at $n_{\mathrm{ph}} = 0.9 \ e/\mathrm{u.c.}$. Initial configurations of the structures are set up from eigenvectors of $M$-point soft phonon modes with different amplitudes and phases (notation explained in the main text).

17

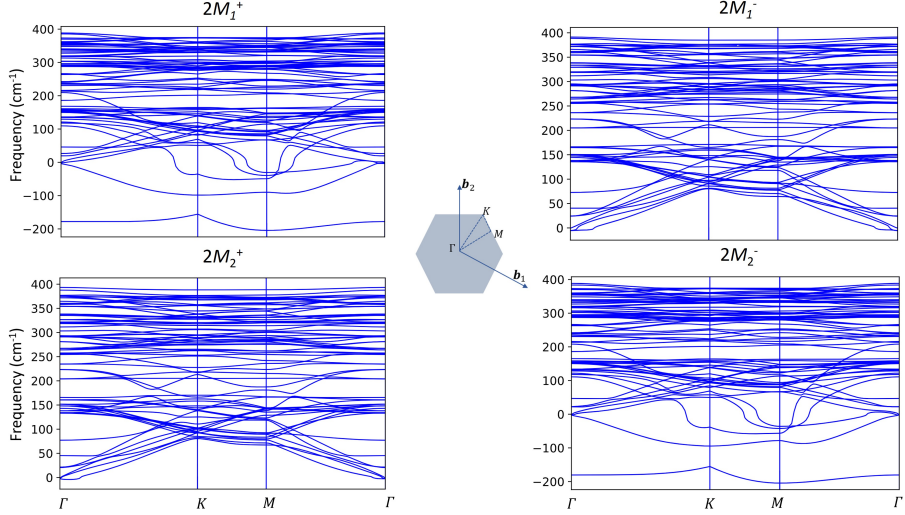

FIG. S16. Similar to Fig. S15, but at $n_{\rm ph} = 1.0$ $e/$u.c..

supercells along the phonon wavevector $\boldsymbol{q}$ direction as:

$$\frac{A}{\sqrt{N_a m_j}} \mathrm{Re}\{e^{i\phi}\boldsymbol{e}_j e^{i\boldsymbol{q}\cdot\boldsymbol{r}_j}\}, \tag{S6}$$

where $A$ and $\phi$ are the input amplitude and phase, $N_a$ is the number of atoms, $m_j$ is the mass of atom $j$, $\boldsymbol{r}_j$ is the position of atom $j$, $\boldsymbol{e}_j$ is the phonon eigenvector on atom $j$. Since at $n_{\rm ph} = 0.9$ $e/$u.c., there are two soft phonon modes at $M$ point denoted as $M_1$ and $M_2$ modes, we set up four sets of initial displacements in the $2\times 2$ reconstructed cell: $2M_1^+$, $2M_2^+$, $2M_1^-$, $2M_2^-$, where "2" denotes the amplitude $A$, and "+/−" denotes a phase $\phi$ of $0°/360°$. We then perform full structural relaxations, followed by phonon calculations on the four well-relaxed structures. It turns out structures relaxed from initial displacements of $2M_1^-$ and $2M_2^+$ have the lowest ground-state energies, which correspond to top-distorted and bottom-distorted structures in Fig. 3 of the main manuscript. The phonon calculations also demonstrate their structural stability, unlike the other two relaxed structures (from initial displacements of $2M_1^+$ and $2M_2^-$) displaying soft phonon modes and thus are unstable. The phonon spectra of these four $2\times 2$ reconstructed structures at $n_{\rm ph} = 0.9, 1.0$ $e/$u.c. are shown in Figs. S15 and S16. In addition, we set up another four sets of initial displacements as the linear combination between $M_1$ and $M_2$ modes, i.e., $2M_1^+ 2M_2^+$, $2M_1^+ 2M_2^-$, $2M_1^- 2M_2^+$ and $2M_1^- 2M_2^-$; the subsequent structural relaxations give the same relaxed configurations as above.

We repeat the same process for soft phonon modes at $K$ point; nevertheless, after struc-

18

| polarization ($10^{-12}$ C/m) | $n_{\rm ph} = 0.9$ $e$/u.c., $2M_1^-$ | $n_{\rm ph} = 0.9$ $e$/u.c., $2M_2^+$ | $n_{\rm ph} = 1.0$ $e$/u.c., $2M_1^-$ | $n_{\rm ph} = 1.0$ $e$/u.c., $2M_2^+$ |
|---|---|---|---|---|
| Total | -4.8 | +5.4 | -4.5 | +5.3 |
| Structural contribution | -1.6 | +2.8 | -2.0 | +3.0 |
| Charge contribution | -3.2 | +2.6 | -2.5 | +2.3 |

TABLE III. Structural and charge contributions to total polarizations of four ground-state structures at $n_{\rm ph} = 0.9, 1.0$ $e$/u.c. .

tural relaxation, atoms move back to original, highly symmetric positions before initial displacements are added. Therefore, no hidden states based on $K$-point soft modes are discovered.

To explore the origin of the giant ferroelectricity of four ground state structures at $n_{\rm ph} = 0.9$ and $1.0$ $e$/u.c., we employed the same method as we used for $n_{\rm ph}$ between $0.1 \sim 0.8$ $e$/u.c.. We computed the polarization for structures with the same configuration as those well relaxed at these two $n_{\rm ph}$ but without any e-h pairs; this part is considered as the structural contribution, and the remaining part is considered as the charge contribution. The results are presented in Table III. Both parts carry significant weights, and that indicates that for these ground-state structures at high number of photoexcited carriers, the polarization should be understood as arising from both origins.

Technologically, our study can further advance the field of optical control of ferroelectric polarization: compared to conventional control using electric field, the optical control does not require complicated circuit setups, and it can significantly improve the control speed from nanosecond to picosecond; this holds great potential for developing the next generation of fast-speed devices[48]. For example, Figure S17(a) below gives an illustrative plot showing the response of bilayer 3R-MoS$_2$ to an ultrafast laser pulse at 2mJ/cm$^2$ (n$_{\rm ph}$ = 0.2 $e$/u.c.). After each light pulse, the polarization decreases from 0.8 pC/m to 0.03 pC/m, resembling a switching from "1" to "0" state in data storage. Moreover, as noted in Ref. 24, a steady-state, high-density electron-hole plasma state can be created with a continuous-wave laser. This suggests that such optical control of polarization can be potentially nonvolatile. This

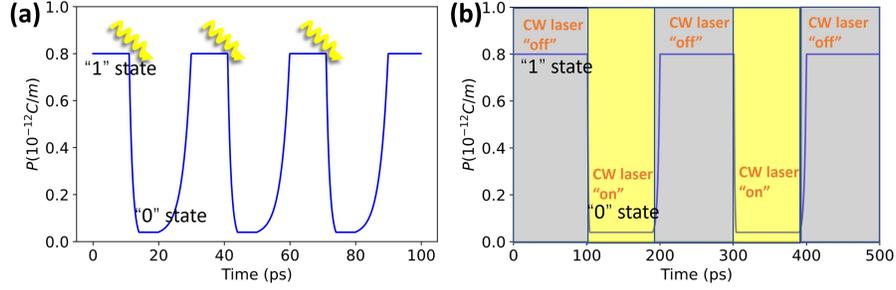

FIG. S17. (a) Illustration about an optically controlled "writing-erasing" process based on bilayer 3R-MoS$_2$. In the normal state, an out-of-plane polarization represents "1" state. With the incidence of an ultrafast laser pulse at 2 mJ/cm$^2$, the polarization decreases to a small value, representing "0" state. Such decrease resembles "erasing" operation. The "0" state recovers to the normal "1" state at a scale of tens of picoseconds[22,50,51], resembling "writing" operation. (b) Nonvolatile "writing-erasing" process when continuous wave laser is the light source. "0" state can be well maintained for a longer time as a steady-state, electron-hole plasma is generated[24].

concept is demonstrated in Fig. S17(b): when light is turned "on" and "off", it corresponds to "erasing" ("0" state) and "writing" data ("1" state), respectively, and it stays in "0" state if we keep light on. Furthermore, when light is at a higher fluence, a transient structural phase transition of bilayer 3R-MoS$_2$ can be induced. Such ultrafast phase change in response to optical signals makes MoS$_2$ a promising phase change material, which can contribute to the development of all-optical devices and neuron-inspired computational technologies[49].



\end{document}